\documentclass[journal]{IEEEtran}
\usepackage{amsmath}
\usepackage{amsfonts}
\usepackage{mathrsfs}
\usepackage{amssymb,color,cite,array}
\usepackage[dvips]{graphicx}

 \allowdisplaybreaks[4]

\begin{document}

\title{ High-Mobility Massive MIMO Communications: Doppler Compensation and Transmit Diversity  }
\author{ Weile Zhang, Zhinan Hu, Shun Zhang, and Gongpu Wang
\thanks{
  W. Zhang and Z. Hu are with the MOE Key Lab for Intelligent Networks and Network Security, Xi'an Jiaotong University, Xi'an, Shaanxi, 710049, China. (email: wlzhang@mail.xjtu.edu.cn)
 }
\thanks{G. Wang is with Beijing Jiaotong University, Beijing 100044, China.}
\thanks{S. Zhang is with the State Key Lab of Integrated Services Networks, Xidian University, Xi'an, Shaanxi, 710071, China (email: zhangshunsdu@gmail.com).}

}

 \maketitle

 \begin{abstract}
 High-mobility  wireless communications have been receiving a lot attentions. In this paper, we consider the high-mobility uplink transmission from a high-speed terminal (HST) to a base station (BS). We propose two transmit diversity schemes for high-mobility massive communications with angle domain Doppler compensation.
 In the first scheme, we propose the idea of block-wise angle domain Doppler compensation, such that the uplink equivalent channel would exhibit the feature of nearly perfect independent block fading. The signal space diversity encoding technique is then exploited to achieve transmit diversity.
In the second scheme, the multiple parallel beamformers with Doppler compensation are exclusively occupied to transmit multiple orthogonal space-time block coding (OSTBC) data streams. Both analysis and numerical results verify that both the proposed schemes can achieve transmit diversity and outperform the conventional scheme.
 \end{abstract}

  \begin{keywords}
High-mobility communications, angle domain massive MIMO,  Doppler compensation, transmit diversity.
 \end{keywords}

\section{Introduction}

High-mobility wireless communications have drawn exploding research
interests from both academic and industry community~\cite{Wu16,HeVTM16}. The relative motion between transceivers lead to Doppler shifts and  thus the fast channel fluctuations. This can pose great challenges on the channel
estimation and equalization.

Basis expansion model (BEM) is a typical approach to realize estimation of fast time-varying channel, where the parameters to be estimated can be significantly reduced~\cite{Wang15,BEM98,MaTVT18}.
The concept of sparse Bayesian learning has been employed to improve the tracking performance of time-varying channels~\cite{BarbuTVT16}.
Another kind of works for high-mobility communications focus on how to benefit system performance by exploiting the Doppler diversity. The pioneering works in this area are performed under the assumption of perfect channel state information (CSI)~\cite{ZhouTWC15}. However, channel estimation errors as well as  channel aging
   are non-negligible and they should have significant impacts on system performance. The Doppler diversity systems with imperfect CSI have been studied in~\cite{ZhouEURASIP15}. The fundamental tradeoff between Doppler diversity and channel estimation errors in high-mobility systems has been quantified for single-input single-output (SISO) and single-input multiple-output (SIMO) systems~\cite{MahamaduAccess18}, respectively.

On the other side, as the Doppler shifts are highly related to multipath angle information, i.e., angle-of-arrival (AoA) for downlink and angle-of-departure (AoD) for uplink, it is shown that, signal processing in angle domain can effectively suppress the time fluctuations of high mobility channels. A few works  have been reported to separate multiple Doppler shifts in angle domain~\cite{ZhangChinaCOM11,YangTB13} for sparse high-mobility  channels.
More recently, large-scale multiple-input multi-output (MIMO) or ``massive MIMO'' has drawn increasing research attentions~\cite{Marzetta10,ZhangTWC18} and exhibits its powerful capability for high-mobility communications especially with richly scatterers due to its high spatial resolution. For example, the work~\cite{Guo17} considered the high-mobility downlink and proposed a Doppler shift estimation scheme by exploiting a high resolution beamforming network. The imperfect calibration of the large scale antenna array has been further considered in~\cite{GeTWC19}. On the other side, for high-mobility uplink transmission,  the angle domain Doppler compensation has been proposed to suppress the harmful effect of Doppler shift with a large-scale antenna array at the transmitter~\cite{Guo19}. The asymptotic scaling law in~\cite{Guo19} showed that the Doppler spread of uplink equivalent channel can  be decreased approximately as $1/\sqrt{M}$ when $M$ is sufficiently large ($M$ is the number of transmit antennas). In~\cite{Ge19}, the exact power spectrum density (PSD) of uplink channel is derived and an antenna weighting technique is then proposed to minimize the channel time variation. Both analysis and numerical results have demonstrated the powerful capability of angle domain Doppler compensation in dealing with time-varying channels~\cite{Guo17,GeTWC19,Guo19,Ge19}. However, the uplink works in~\cite{Guo19,Ge19} only considered a single transmit data  stream and may suffer from the loss of transmit diversity gain.

In this paper, we propose two transmit diversity schemes for high-mobility massive communications with angle domain Doppler compensation. To the best of our knowledge, this work is the first to exploit the transmit diversity with angle domain Doppler compensation for high-mobility communications.  
   In the first proposed scheme, we propose the idea of block-wise angle domain Doppler compensation, which independently re-generate the random phases for the beamformers in each block. By doing so, the uplink equivalent channel would exhibit the feature of nearly perfect independent block fading. The signal space diversity (SSD) encoding technique is then exploited to achieve transmit diversity.
In the second proposed scheme, the multiple parallel beamformers with Doppler compensation are exclusively occupied to transmit multiple multiple orthogonal space-time block coding (OSTBC) data streams. Both analysis and numerical results show that both the proposed schemes can achieve transmit diversity.


\section{Angle-Domain Doppler Compensation}

In this section, we first briefly review the concept of angle domain Doppler compensation for high-mobility uplink transmissions~\cite{Guo19,Ge19}.
As illustrated in Fig. 1, we consider the high-mobility uplink transmission from a high-speed train (HST) to a base station (BS), where an $M$-element uniform linear array (ULA) is equipped at HST. We assume BS has only one antenna.
 Denote the normalized antenna space of ULA as $d$, which is the ratio between the antenna spacing and the radio wavelength. Let the direction of ULA coincide with that of HST motion.
We consider the flat fading channel for simplicity.
Nevertheless, the proposed scheme can be directly applied to frequency selective channels.
Specifically, the channel from HST to BS is composed of a bunch of propagation paths. We consider the typical Jake's model such that the angle of departure (AOD) of these paths be constrained within a region of $(0, \pi)$.

\begin{figure}[t]
\centering
\includegraphics[width=9cm]{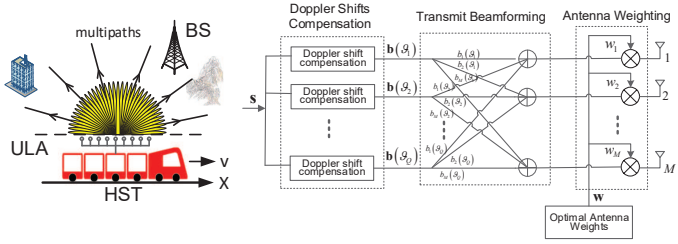}
\vspace{-5mm}
\caption{Uplink transmission with angle-domain Doppler compensation from a HST to a BS.}
\end{figure}

Denote ${{\bf s}} = [{s}(0),{s}(1),...,{s}(N - 1)]^T$ as the length-$N$ transmitted time domain symbols in one orthogonal frequency division multiplexing (OFDM) block. We consider the antenna weighting technique~\cite{Ge19} is adopted, where the $m\rm{th}$ antenna is weighted by a complex weight $w_m$.
 The multi-branch transmit matched filtering (MF) beamforming is performed towards a set of $Q$ selected directions
${{\vartheta }_{q}}\in(0 , \pi),q=1,2,...,Q$.
The $q$th MF beamformer is
\begin{align}
{\bf b}({\vartheta _q}) = \zeta \cdot{\rm{\bf a}(}{\vartheta _q}){e^{j\pi \phi_q }},
 \end{align}
 where ${\bf a}(\vartheta_q) = [1, {\rm e}^{j2\pi d \cos\vartheta_q}, \cdots, {\rm e}^{j2\pi d (M-1)\cos\vartheta_q}]^T$ denotes the array response vector corresponding to direction $\vartheta_q$,  $\zeta$ is the normalization coefficient to keep the total transmit power per symbol to 1 and $\phi_q $ denotes the introduced random phase.

Let ${\bf\Phi}(  \epsilon ) = \textrm{diag}(1, {\rm e}^{  j \omega_d  T_s \epsilon }, \cdots, {\rm e}^{  j \omega_d T_s (N-1)\epsilon } )$ represent the $N\times N$ diagonal phase rotation matrix introduced by frequency shift of $ f_d \epsilon$,  with $\omega_d = 2\pi f_d$  and $T_s$ being sampling interval.
After transmit beamforming and Doppler shift compensation, the transmitted $M\times N$ signal matrix at transmit antenna array can be expressed as $ {\bf X}_q =  \textrm{diag}( {\bf w}) {\bf b}^*({\vartheta _q}) {\bf s}^T {\bf\Phi}(  -\cos\vartheta_q  )  $.
Then, according to~\cite{Guo19,Ge19}, the received signal at BS can be expressed in the form of
\begin{align}\label{equ1}
 {\bf r} = &  \sum_{q=1}^Q  \int_{0}^{\pi}  \alpha(\theta)
{\bf a}(\theta)^T {\bf X}_q {\bf\Phi}(  \cos\theta  )    d\theta,
 \nonumber \\
 = &  \sum_{q=1}^Q  \int_{0}^{\pi}  \alpha(\theta)
{\bf a}(\theta)^T \textrm{diag}( {\bf w})  {\bf b}^*({\vartheta _q}) {\bf s}^T \nonumber \\ & \kern 80pt \times {\bf\Phi}(  -\cos\vartheta_q  ) {\bf\Phi}(  \cos\theta  )    d\theta,
\end{align}
where $\alpha(\theta)$ represents the complex gain of the path associated with AOD $\theta$ and ${\bf w} = [w_1, w_2, \cdots, w_M]^T \in\mathbb{C}^{M\times 1}$ denotes the weighting vector.

Given sufficiently large $M$, we can make the approximation: ${\bf a}(\theta)^T \textrm{diag}( {\bf w})  {\bf a}^*({\vartheta _q}) \simeq 0$ with $\theta \ne \vartheta_q$. We can then rewrite (\ref{equ1}) into
\begin{align}\label{equ2}
 {\bf r} =  \underbrace{ \left( \zeta \cdot \big( \sum_m w_m \big)  \sum_{q=1}^Q    \alpha(\vartheta_q )
   {\rm e}^{-j\pi \phi_q} \right) }_{h_{\rm eq}} {\bf s}^T  =  h_{\rm eq}\cdot {\bf s}^T .
\end{align}
It is seen that, with sufficiently large $M$, the equivalent channel at BS can be considered as time invariant. The  uplink equivalent channel can be expressed as $h_{\rm eq}$ as defined in (\ref{equ1}).

\section{SSD based Transmit Diversity Scheme with Doppler Compensation}

Assume that the information bits are first mapped to a block of $N$ data symbols, ${\bf d} = [d(1), d(2), \cdots, d(N)]^T$.  Assume $N = K\times J$ and rewrite ${\bf d}$ into $J$ length-$K$ subvectors, i.e., ${\bf d} = [{\bf d}(1)^T, {\bf d}(2)^T, \cdots, {\bf d}(J)^T  ]^T$ with ${\bf d}(j)\in \mathbb{C}^{K\times 1}$.

\begin{figure}[t]
\centering
\includegraphics[width=5cm]{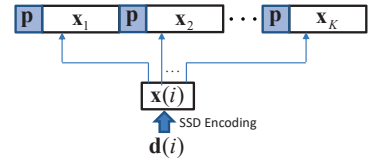}
\vspace{-3mm}
\caption{Illustration of the transmission frame of  transmit diversity scheme based on SSD.}
\end{figure}

Denote $\tilde{K} = 2^{\lceil \log_2 K\rceil}$ and $\tilde{\bf \Theta} = {\bf F}^H_{\tilde{K}} \textrm{diag}(1, {\rm e}^{{\bf j}\pi/(2\tilde{K})},  \cdots $,  ${\rm e}^{{\bf j}\pi(\tilde{K}-1)/(2\tilde{K})}  )$, where ${\bf F}^H_{\tilde{K}}$ is the normalized $\tilde{K}\times \tilde{K}$ discrete Fourier transform (DFT) matrix.
Assume each subvector ${\bf d}(j)$ is individually encoded as:
\begin{align}\label{equSSD}
{\bf x}(i) = {\bf \Theta} {\bf d}(i)
\end{align}
where ${\bf \Theta} \in\mathbb{C}^{K\times K}$ is the first principal $K\times K$ submatrix of $\tilde{\bf\Theta}$. The encoding in (\ref{equSSD}) is referred to as the signal space diversity technique~\cite{Su05}, in which the original signal constellation is rotated and expanded such that zero entry in the vector ${\bf x}(i)$ does not exist for any $ {\bf d}(i) \ne {\bf 0}$.

Rewrite ${\bf x}(i) $ as ${\bf x}(i) = [ x_1(i), x_2(i), \cdots, x_K(i) ]^T$.
Then, the $k\rm{th}$ data block can be formed as
\begin{align}
{\bf x}_k = [ x_k(1), x_k(2), \cdots, x_k(J) ]^T.
\end{align}

The $k\rm{th}$ transmitted signal block is composed of length-$Q$ pilot block ${\bf p} \in\mathbb{C}^{N_p\times 1}$ with $N_p$ known pilot symbols and the data block ${\bf x}_k$ as follows:
\begin{align}
{\bf s}_k = [ {\bf p}^T,  \quad {\bf x}_k^T ]^T.
\end{align}

In the proposed scheme,  the signal blocks ${\bf s}_k$, $k=1,2,\cdots, K$ are transmitted with angle-domain Doppler shift compensation.
However, different from the conventional scheme which employs one realization of the random phases for the beamformers during the whole transmission frame~\cite{Guo19,Ge19},
 we propose to independently re-generate the random phases for the beamformers in each signal block. Mathematically, the $q\rm{th}$ MF beamformer in the $k\rm{th}$ data block can be expressed as
\begin{align}
{\bf b}_k({\vartheta _q}) = \zeta \cdot{\rm{\bf a}(}{\vartheta _q}){e^{j\pi \phi_{k,q} }},
 \end{align}
where $\phi_{k,q}$ denotes the introduced random phase re-generated for the $k\rm{th}$ block.

Then,  following the similar steps as (\ref{equ1}) to (\ref{equ2}), with sufficiently large $M$, the received signal in the noise-free situation corresponding to the $k\rm{th}$ block can be expressed as
 \begin{align}
 {\bf r}_k =  \underbrace{ \left( \zeta \cdot \big( \sum_m w_m \big)   \sum_{q=1}^Q    \alpha(\vartheta_q )
   {\rm e}^{-j\pi \phi_{k,q}} \right) }_{h_{k, {\rm eq} }} {\bf s}^T_k  =  h_{k, {\rm eq}}\cdot {\bf s}_k^T .
\end{align}
where $ h_{k, {\rm eq}}$ represents the equivalent time-invariant channel in the $k\rm{th}$ block. Note that the equivalent invariant channel parameters $ h_{k, {\rm eq}}$, $k=1,2,\cdots,K$, can be directly estimated from the pilot blocks.

It is evident that, owing to the introduced random phases individually for each block, the
uplink equivalent channels of different blocks are independent with each other, or mathematically, there holds $E[h_{k, {\rm eq}} h_{k',{\rm eq}}^*] = 0$ for any $k\ne k'$.
Hence, in the asymptotic case, i.e., sufficiently large $M$,  it is observed that
the uplink equivalent channel would exhibit the feature of  \textit{perfect independent block fading} in the proposed scheme. That is, uplink equivalent channel $h_{k, {\rm eq}}$ holds constant within each block but varies independently across the blocks. Thus, the SSD encoding procedure is expected to be able to achieve transmit diversity.

Let us collect the received symbols corresponding to the $i\rm{th}$ encoded block ${\bf x}(i)$, that yields
 \begin{align}\label{equ3}
 {\bf y}(i) = & \Big[ {\bf r}_1(N_p+i), {\bf r}_2(N_p+i), \cdots, {\bf r}_K(N_p+i)  \Big]^T \nonumber \\ = &  {\bf H}_{\rm eq} {\bf\Theta} {\bf d}(i),
 \end{align}
 where $ {\bf H}_{\rm eq} \in\mathbb{C}^{K\times K}$ can be expressed as
 \begin{align}
 {\bf H}_{\rm eq} = \textrm{diag}(h_{1, {\rm eq}}, h_{2, {\rm eq}}, \cdots, h_{K, {\rm eq}} ).
 \end{align}
 Then, the maximum likelihood (ML) data detection can be performed as
 \begin{align}\label{equ4}
\hat{\bf d}(i) =  \arg\min_{\forall \tilde{\bf d}(i)} \| {\bf y}(i) - {\bf H}_{\rm eq} {\bf\Theta} \tilde{\bf d}(i) \|.
 \end{align}
 It is observed that the data symbol vector ${\bf d}(i)$ has been encoded and spread over $K$ independent fading channels to achieve the transmit diversity. According to~\cite{Su05}, it is expected a diversity order of $K$ can be achieved in the proposed scheme.

\section{OSTBC based Transmit Diversity Scheme with Doppler Compensation}

In this section, we present the proposed OSTBC based transmit diversity scheme with Doppler compensation. The basic idea of the proposed scheme is  assign  multiple OSTBC encoded data streams exclusively to the multiple parallel beamformers with Doppler compensation.  For brevity, we take the Alamouti coding for example.
Nevertheless, the proposed scheme can be directly extended for higher OSTBC structure.

\begin{figure}[t]
\centering
\includegraphics[width=6cm]{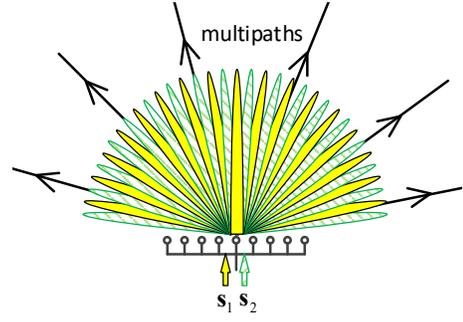}
\vspace{-3mm}
\caption{Illustration of Alamouti based transmit diversity scheme with Doppler compensation.}
\end{figure}

Assume that the information bits are first mapped to a block of $N$ data symbols, ${\bf d} = [d(1), d(2), \cdots, d(N)]^T$. Consider ${\bf d}$ can be divided into two length-$N/2$ vectors: ${\bf d} = [{\bf x}_1^T, {\bf x}_2^T ]^T \in\mathbb{C}^{N\times 1}$ with ${\bf x}_i \in \mathbb{C}^{N/2\times 1}$.
The Almouti coded transmission frame can be expressed as
\begin{align}
\left[ \begin{array}{c}{\bf s}_1^T \\ {\bf s}_2^T  \end{array}\right] = \left[
\begin{array}{cccc}
{\bf p}^T, & {\bf 0}_{1\times L}, & {\bf x}_1^T, & -{\bf x}_2^H \\
 {\bf 0}_{1\times L}, & {\bf p}^T, & {\bf x}_2^T, & {\bf x}_1^H
\end{array}
\right].
\end{align}
Here, ${\bf s}_1$ and ${\bf s}_2$ denote the two encoded signal streams, which are initialized with time division pilot block ${\bf p}\in \mathbb{C}^{N_p\times 1}$.

Denote $Q_{odd}$ as the set of odd indices from 1 to $Q$, and $Q_{even}$ as the set of even indices from 1 to $Q$. Among all the $Q$ beamformers, let the odd beamformers transmit the signal vector ${\bf s}_1$, while the even beamformers transmit the signal vector ${\bf s}_2$. Mathematically, the transmitted $M\times 2(N+L)$ signal matrix for the $q\rm{th}$ beamformer can be expressed as
\begin{align}
{\bf X}_q = \left\{
\begin{array}{ll}
 \textrm{diag}( {\bf w}) {\bf b}^*({\vartheta _q}) {\bf s}_1^T {\bf\Phi}(  -\cos\vartheta_q  ), & q  \in Q_{odd}, \\
 \textrm{diag}( {\bf w}) {\bf b}^*({\vartheta _q}) {\bf s}_2^T {\bf\Phi}(  -\cos\vartheta_q  ), & q  \in Q_{even}.
\end{array}
  \right.
\end{align}

In the noise-free situation, the received signal at BS can be expressed as
\begin{align}
 {\bf r} = &  \sum_{q \in Q_{odd}}  \int_{0}^{\pi}  \alpha(\theta)
{\bf a}(\theta)^T \textrm{diag}( {\bf w})  {\bf b}^*({\vartheta _q}) {\bf s}_1^T{\bf\Phi}(  -\cos\vartheta_q  )   \nonumber \\ &   \times {\bf\Phi}(  \cos\theta  )    d\theta  + \sum_{q \in Q_{even}}  \int_{0}^{\pi}  \alpha(\theta)
{\bf a}(\theta)^T \textrm{diag}( {\bf w})  {\bf b}^*({\vartheta _q}) {\bf s}_2^T \nonumber \\ &  \times {\bf\Phi}(  -\cos\vartheta_q  ) {\bf\Phi}(  \cos\theta  ).
\end{align}

Given sufficiently large $M$, we obtain:
\begin{align}
 {\bf r} = &  \underbrace{ \left( \zeta \cdot  \big( \sum_m w_m \big)   \sum_{q\in Q_{odd}}    \alpha(\vartheta_q )
   {\rm e}^{-j\pi \phi_q} \right) }_{h_{1, {\rm eq}}} {\bf s}_1^T \nonumber \\
& \kern 20pt + \underbrace{ \left( \zeta \cdot   \big( \sum_m w_m \big) \sum_{q\in Q_{even}}    \alpha(\vartheta_q )
   {\rm e}^{-j\pi \phi_q} \right) }_{h_{2, {\rm eq}}} {\bf s}_2^T \nonumber \\
   = & h_{1, {\rm eq}} {\bf s}_1^T   + h_{2, {\rm eq}} {\bf s}_2^T.
\end{align}

It is evident that, with sufficiently large $M$, i.e., the asymptotic case, the uplink equivalent channels of the two signal streams can be considered time invariant. As a consequence, BS can directly acquire the channel estimation for $h_{1, {\rm eq}}$ and $h_{2, {\rm eq}}$  based on the two time division pilot blocks, and then perform data detection based on the Alamouti coding structure. It is expected a diversity order of two can be achieved.


\section{Simulations}

In this section, we present simulation results to verify the proposed studies. We consider a ULA with a total of $M=64$ antennas and normalized antenna spacing of 0.45. The symbol rate is assumed as 1 MHz. The radio frequency is taken as 5.5 GHz. The moving speed is taken as 100 m/s. The other parameters are set as: $N_p=16$ and $J=64$. QPSK constellation is adopted.

We evaluate the symbol error rate (SER) performance of the proposed scheme as a function signal-to-noise ratio (SNR) in Fig. 4. The proposed SSD and Alamouti based transmit diversity schemes with Doppler compensation are labelled as `SSD-DC' and `Alamouti-DC', respectively. The convention Doppler compensation transmit scheme~\cite{Ge19} without diversity is referred to as `No-Diversity-DC'. We consider that No-Diversity-DC  employs exactly the same frame structure as Fig. 2, with ${\bf x}_k$, $k=1,2,\cdots,K$, being data symbols directly drawn from QPSK constellations.  The following observations can be made:

First, the performance superiority of the proposed schemes is evident. Especially, from the slopes of the curves, it is seen that the SER of the proposed diversity schemes would drops more quickly as compared to No-Diversity-DC. This should be attributed to the exploited transmit diversity in the proposed schemes.

Second, we also see that in the region with sufficient SNR,  SER of the proposed SSD-DC with $K=2$ and Alamouti-DC can be reduced in an order of two when SNR is increased by 10 dB.
This coincides with the previous discussions that a diversity order of two can be achieved in these situations.

Third, we can observe a little performance improvement of SSD-DC as compared to Alamouti-DC. We can also see the diversity improvement of SSD-DC when $K$ is increased from 2 to 4.
Nevertheless, we should note that, in order to achieve diversity, SSD-DC scheme relies on ML detection of joint $K$ data symbols, whereas symbol-wise detection is sufficient in Alamouti-DC scheme.
Hence, there exists the tradeoff between detection performance and computational complexity in the proposed schemes.

\begin{figure}[t]
\centering
\includegraphics[width=8cm]{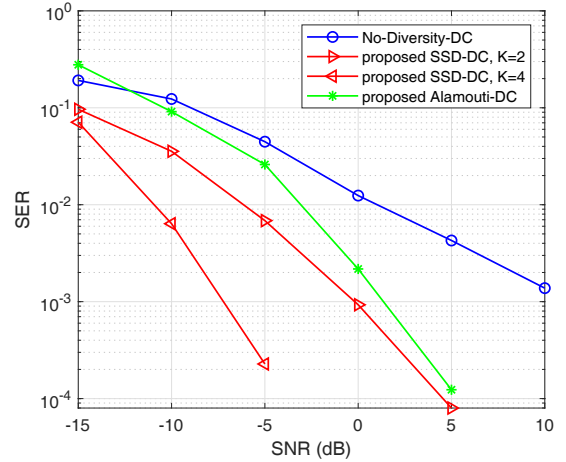}
\vspace{-3mm}
\caption{SER performance comparison between the conventional scheme and the proposed diversity schemes.}
\end{figure}

\section{Conclusions}

In this paper, we proposed two transmit diversity schemes for high-mobility massive communications with angle domain Doppler compensation based on SSD and OSTBC, respectively. Both analysis and numerical results showed that both the proposed schemes can achieve transmit diversity can substantially outperform the conventional scheme.

\end{document}